\documentclass{article}
\usepackage{latexsym}
\newtheorem{corollary}{Corollary}

\newtheorem{prop}{Proposition}
\newtheorem{definition}{Definition}
\newtheorem{lemma}{Lemma}
\newtheorem{theorem}{Theorem}
\begin{document}

\title{Cones of ball-ball separable elements}

\author{Roland Hildebrand \thanks{%
LMC, Universit\'e Joseph Fourier, Tour IRMA, 51 rue des
Math\'ematiques, 38400 St.\ Martin d'H\`eres, France ({\tt
roland.hildebrand@imag.fr}). This paper presents research results
of the Action Concert\'ee Incitative "Masses de donn\'ees" of CNRS,
France. The scientific responsibility rests with its author.}}

\maketitle

\begin{abstract}
Let $B_1,B_2$ be balls in finite-dimensional real vector spaces $E_1,E_2$, centered around unit length
vectors $v_1,v_2$ and not containing zero. An element in the tensor product space $E_1 \otimes E_2$ is called
$B_1 \otimes B_2$-separable if it is contained in the convex conic hull of elements of the form $w_1 \otimes
w_2$, where $w_1 \in B_1$, $w_2 \in B_2$. We study the cone formed by the separable elements in $E_1 \otimes
E_2$. We determine the largest faces of this cone via a description of the extreme rays of the dual cone,
i.e.\ the cone of the corresponding positive linear maps. We compute the radius of the largest ball centered
around $v_1 \otimes v_2$ that consists of separable elements. As an application we obtain lower bounds on the
radius of the largest ball of separable unnormalized states around the identity matrix for a multi-qubit system. These
bounds are approximately 12\% better than the best previously known. Our results are extendible to the case
where $B_1,B_2$ are solid ellipsoids.
\end{abstract}

\section{Introduction}

Let $K,K'$ be closed convex pointed cones with non-empty interior, residing in finite-dimensional real
vector spaces $E,E'$. Then an element $w \in E \otimes E'$ of the tensor product space is called {\sl separable}
if it can be represented as a convex combination of product elements $v \otimes v'$, where $v \in K$, $v' \in K'$.
It is not hard to show that the set of separable elements is itself a closed convex pointed cone with non-empty
interior. This cone is called the $K \otimes K'$-separable cone. Cones of elements that are separable with respect
to more than two initial cones are defined similarly.

Separable cones have many applications in Mathematical Programming. So, the dual cones to the cones of
positive polynomials, which frequently appear in optimization problems \cite{PosPolsinControl},\cite{NesterovSOS}, can be
represented as separable cones. In Quantum Information Theory the set of unnormalized unentangled mixed
states of a multi-partite quantum system also forms a separable cone. In this case the underlying cones are
cones of positive semidefinite matrices. These separable cones and the corresponding positive maps have been
subject of intense study in the recent Quantum Information Theory literature \cite{Gurvits0302102},\cite{Horodeckis96},\cite{Peres96},
but drew attention of the mathematical community also before the emergence of this field \cite{Choi74},\cite{Stormer63},\cite{Terpstra},\cite{Woronowicz}.

In this paper we treat ball-ball separable cones, i.e.\ cones of $K \otimes K'$-separable elements where the
underlying cones $K,K'$ are conic hulls of closed Euclidean balls or solid ellipsoids not containing the
origin. Such cones have a relatively simple structure. Thus they are suitable for the approximation of more
complex separable cones. This can be done by approximating the underlying cones by appropriate ball-generated
cones. The idea of replacing an underlying cone by a ball-generated cone was put forward by Leonid Gurvits and Howard Barnum, who
successfully used it to obtain lower bounds on the largest ball of unnormalized separable elements around
the identity matrix for multipartite systems \cite{Gurvits0302102},\cite{Gurvits0409095}. In this contribution we compute several characteristics of
ball-ball separable cones exactly, which allows for a more efficient application of such approximations.

One such application makes use of the fact that the cone of positive semidefinite hermitian $2 \times 2$
matrices is isomorphic to the 4-dimensional Lorentz cone $L_4$ and hence is also a ball-generated cone. This
enables us to refine Gurvits' bounds for the case of multi-qubit systems. We prove that for a 3-qubit system
a ball of radius $\sqrt{4/5}$ around the identity matrix consists only of separable elements, as opposed to the best bound
$\sqrt{8/11}$ known previously \cite{Gurvits0409095}. For systems consisting of more than 3 qubits we obtain an
improvement of roughly 12\% with respect to the best bounds known before \cite{Gurvits0409095}. Namely, we prove
that for an $m$-qubit system, a ball of radius $\frac{2^{m/2}}{\sqrt{3^{m-1}+1}}$ around the identity matrix consists only
of separable elements. The exponent in the asymptotics (as $m \to \infty$) of this bound is the same as the
one obtained in \cite{Gurvits0409095}. Recently Stanislaw Szarek showed that this exponent delivers the exact asymptotics
in the multi-qubit case (an earlier result is published in \cite{Szarek0310061}).

\smallskip

The paper is organized as follows.

In the next section we characterize the extreme rays of the cone dual to the cone of ball-ball separable
elements, namely the cone of linear maps that take the Lorentz cone to the Lorentz cone in the respective
source and target spaces (Lorentz-to-Lorentz positive maps). It is well-known that the extreme rays of the
dual cones characterize the largest faces in the primal cones \cite{Rockafellar}. These faces are of interest to
us because they determine the radii of the largest separable balls around chosen elements in the separable
cones. Note that describing the extreme rays of cones of Lorentz-to-Lorentz positive maps yields also a
description of cones dual to ellipsoid-ellipsoid separable cones, because extreme rays are taken to extreme
rays by invertible linear mappings.

In Section 3 we describe the largest faces of ball-ball separable cones using the obtained families of
extreme rays in the dual cones. This allows to get some insight into the structure of ball-ball separable
cones.

In Section 4 we compute the radius of the largest ball of ellipsoid-ellipsoid separable elements around the tensor
product of points defining the central rays of the initial ellipsoid-generated cones.

In Section 5 we apply this result to study the cone of separable unnormalized states of a multi-qubit system. We compute the above-mentioned lower bounds
on the radii of largest separable balls around the unnormalized uniformly mixed state.

In the last section we summarize our results.

\section{Extreme rays of cones of Lorentz-to-Lorentz positive maps}

In this section we compute the extreme rays of the cone of positive maps, i.e.\ those linear maps which take the Lorentz cone in the Lorentz cone
in the respective source and target spaces.

\smallskip

Let $L_m \subset {\bf R}^m$, $L_n \subset {\bf R}^n$ be standard Lorentz cones of dimensions $m$ and $n$, i.e.
\begin{eqnarray*}
L_m &=& \{(x_0,x_1,\dots,x_{m-1})^T \,|\, x_0 \geq
|(x_1,\dots,x_{m-1})^T|\}, \\ L_n &=& \{(y_0,y_1,\dots,y_{n-1})^T
\,|\, y_0 \geq |(y_1,\dots,y_{n-1})^T|\}.
\end{eqnarray*}
We assume throughout the paper that $\min(n,m) \geq 2$. Since $L_1$ is isomorphic to the ray ${\bf R}_+$, the
case $\min(n,m) = 1$ is trivial. We call a linear map $M: {\bf R}^m \to {\bf R}^n$ {\sl $L_m$-to-$L_n$
positive} or just {\sl positive} if $M[L_m] \subset L_n$. Since the Lorentz cones are self-dual, the cone
${\cal P}$ of such maps is dual to the cone of $L_m \otimes L_n$-separable elements. Moreover, as a
consequence of this self-duality $M$ is positive if and only if the adjoint map $M^T$ is positive in the
sense that $M^T[L_n] \subset L_m$.

In this section we determine the extreme rays of the cone ${\cal
P}$ of positive maps. We represent maps from ${\bf R}^m$ to ${\bf
R}^n$ by $n \times m$ matrices partitioned as
\begin{equation} \label{partition}
M = \left( \begin{array}{cc} s & h \\ v & A \end{array}
\right),
\end{equation}
where $s$ is a scalar, $h$ is a row vector, $v$ is a column vector
and $A$ is a $(n-1)\times(m-1)$-matrix. Note that if $M$ is a
non-zero positive map, then the scalar $s$ is strictly positive.

Define two diagonal matrices $J_n = diag(1,I_{n-1})$, $J_m =
diag(1,I_{m-1})$, where $I_k$ denotes the $k \times k$ identity matrix. Note that if $x \in {\bf R}^m$ is in the
interior of $L_m$, then $x^TJ_mx > 0$. If $x \in \partial L_m$,
then $x^TJ_mx = 0$.

{\lemma \label{poscond} A map
\[ M = \left( \begin{array}{cc} 1 & h \\ v & A \end{array} \right) \in {\bf R}^{n \times m}
\]
is $L_m$-to-$L_n$ positive if and only if
\[ \exists\, \lambda \geq 0: \quad M^T J_n M \succeq \lambda J_m, \quad |h| \leq 1,
\]
or equivalently,
\[ \exists\, \lambda' \geq 0: \quad M J_m M^T \succeq \lambda J_n, \quad |v| \leq 1.
\] }

{\it Proof.} By definition, $M$ is positive if for any $x_0 \geq
0$, $x \in {\bf R}^{m-1}$ such that $x_0 \geq |x|$ we have $y_0
\geq |y|$, where
\[ \left( \begin{array}{c} y_0 \\ y \end{array} \right) = M \left( \begin{array}{c} x_0 \\ x \end{array}
\right).
\]
We can rewrite this equivalently as
\[ \forall\,x_0,x\ |\ x_0 \geq 0,\ ( x_0\ x^T) J_m \left( \begin{array}{c} x_0 \\ x \end{array} \right) \geq 0: \quad
x_0 + hx \geq 0, \quad ( x_0\ x^T) M^T J_n M \left( \begin{array}{c} x_0 \\ x \end{array} \right) \geq 0.
\]
By the ${\cal S}$-lemma \cite{Dines43},\cite{Yakubovich71} this is equivalent to the conditions
\[ (1\ h)^T \in L_m, \quad \exists\, \lambda \geq 0: \quad M^T J_n M \succeq \lambda J_m,
\]
which gives the first set of conditions claimed by the lemma. The
second set is obtained by considering the adjoint maps $M^T$ in
the $L_n$-to-$L_m$ positive cone. $\Box$

\smallskip

Let us establish necessary and sufficient conditions for a
positive map to generate an extreme ray of the cone ${\cal P}$. Note also that $M$ generates an extreme ray if and only if $M^T$ generates an extreme ray.
First we show that if a non-zero positive map $M$ does not take
the interior of $L_m$ in the interior of $L_n$, then $M$ is of
rank 1.

{\lemma \label{rk1} Let $M \not= 0$ be a positive map. Suppose there exists $x
\in int\,L_m$ such that $Mx \in \partial L_n$. Then the rank of
$M$ is equal to 1. }

{\it Proof.} Let the conditions of the lemma hold. Denote the point $Mx \in \partial L_n$ by $y$. Then $y$ is
contained in the linear subspace $M[{\bf R}^m] = Im\,M$. Since $M \not= 0$, this subspace is non-zero.
Moreover, since $x \in int\,L_m$, there exists a neighbourhood $U_m \subset {\bf R}^m$ of $x$ which is
entirely contained in $L_m$. Its image $M[U_m]$ will be a neighbourhood $U_n$ of $y$ relative to $Im\,M$. By
the positivity of $M$ the set $U_n$ is contained in $L_n$ and therefore in the face of $y$ with respect to
the cone $L_n$. But $y \in \partial L_n$, hence this face equals the intersection of $L_n$ with the linear
subspace generated by $y$. Therefore $y \not= 0$ and $Im\,M = \{ \alpha y\,|\, \alpha \in {\bf R}\}$. Thus
$M$ has rank 1. $\Box$

{\lemma \label{rk1extr} A positive map $M$ of rank 1, partitioned as in (\ref{partition}), generates an
extreme ray if and only if $|h| = |v| = s$. }

{\it Proof.} Let $M$ be a rank 1 map, partitioned as in
(\ref{partition}), and let $s > 0$. Then we have
\[ M = \frac{1}{s} \left( \begin{array}{c} s \\ v \end{array}
\right) \left( s \ h \right).
\]
The condition of positivity provided by Lemma \ref{poscond} then
takes the form
\[ (s\ h)^T \in L_m, \quad \exists\, \lambda \geq 0: \quad \frac{1}{s^2} \left( \begin{array}{c} s \\ h^T \end{array}
\right) \left( s \ v^T \right) J_n \left( \begin{array}{c} s \\
v \end{array} \right) \left( s \ h \right) \succeq \lambda J_m
\]
\[ \Leftrightarrow\ s \geq |h|,\ \exists\lambda \geq 0:\ \frac{1}{s^2} (s^2 -
|v|^2) \left( \begin{array}{c} s \\ h^T \end{array} \right) \left(
s \ h \right) \succeq \lambda J_m
\]
\begin{equation} \label{poscrk1}
\Leftrightarrow\ s \geq |h|,\ s \geq |v|.
\end{equation}
Hence a rank 1 map is positive if and only if $s \geq |h|$, $s
\geq |v|$, $s > 0$.

Let now $M$ be a positive map of rank 1 and suppose that $s >
|h| \not= 0$. Then the maps
\[ M(\lambda) = \frac{1}{s} \left( \begin{array}{c} s \\ v \end{array}
\right) \left( s \ \lambda h \right)
\]
are positive for all $\lambda \in [-s/|h|,s/|h|]$ and $M = M(1)$. The point $\lambda = 1$ lies in the interior of the interval $[-s/|h|,s/|h|]$,
and the matrices $M(\lambda)$ are not multiples
of each other for different $\lambda$. Hence $M$ does not generate an extreme ray of ${\cal P}$.

A slightly modified argument can be applied if $h = 0$. Choose any
vector $h'$ with $|h'| = s$ and consider the maps
\[ M(\lambda) = \frac{1}{s} \left( \begin{array}{c} s \\ v \end{array}
\right) \left( s \ \lambda h' \right).
\]
Then $M(\lambda)$ is positive for all $\lambda \in [-1,1]$ and $M = M(0)$. Hence $M$ cannot
generate an extreme ray neither.

The same reasoning applies if $s > |v|$. Thus if $M$ generates an extreme ray, then $|h| = |v| = s$.

It rests to show that any rank 1 matrix with $|h| = |v| = s > 0$
generates an extreme ray. Let $M$ be such a matrix. Suppose there
exists a matrix $\delta M$ such that $M(\lambda) = M + \lambda
\delta M \in {\cal P}$ for all $\lambda$ in a neighbourhood $U$ of
zero. Let $V$ be a neighbourhood of the unit vector $e_0 \in {\bf
R}^m$ that lies entirely in the interior of $L_m$. Then for all
$\lambda \in U$ and for all $w \in V$ we have
\[ M(\lambda) w = Mw + \lambda\,\delta M\,w \in L_n,
\quad Mw = \frac{1}{s} (s\ h)w \cdot \left( \begin{array}{c} s \\ v \end{array} \right) \in \partial L_n.
\]
Since $\lambda$ can vary in a neighbourhood of zero, the vectors $Mw + \lambda\,\delta M\,w$ have to lie in
the face of $Mw$ with respect to the cone $L_n$. Then $\delta M\,w$ lies in the tangent space to that face.
This tangent space is the linear subspace generated by $Mw$. Hence $\delta M\,w$ has to be a multiple of the
vector $Mw$. Since this holds for all $w \in V$, the image of $\delta M$ must be contained in the linear
subspace of ${\bf R}^n$ generated by the set $\{Mw\,|\,w \in V\}$. Note that $\frac{1}{s} (s\ h)w > 0$,
because $w \in int\,L_m$. Therefore this subspace is one-dimensional and generated by $(s\ v^T)^T$. It
follows that $\delta M$ is of the form $(s\ v^T)^T u$ for some vector $u \in {\bf R}^m$. If we apply the same
line of reasoning for the positive map $M^T$, we conclude that $\delta M^T$ is of the form $(s\ h)^T u'$ for
some vector $u' \in {\bf R}^n$. Thus $\delta M$ is proportional to $M$ and $M$ generates an extreme ray of
${\cal P}$.

This completes the proof of the lemma. $\Box$

\smallskip

It rests to consider the positive maps of rank strictly greater than 1. Let $M$ be such a map, partitioned as in (\ref{partition}). By Lemma \ref{rk1} $M$ takes the interior of $L_n$ to the interior of $L_m$.
Let $Aut(L_m)$, $Aut(L_n)$ be the automorphism groups of the cones $L_m$, $L_n$, respectively.
We shall now show that if $M$ generates an extreme ray, then there exist automorphisms $U_n \in Aut(L_n)$, $U_m \in Aut(L_m)$ such that
$U_nMU_m$ is {\sl doubly stochastic}. (A positive map $M$ is called doubly stochastic if
$M$ and $M^T$ take the central elements $e_0$ of the cones $L_m,L_n$ into each other. Otherwise spoken, $M$ is
doubly stochastic if $s = 1$ and $h = v = 0$.)

Define two functions $p,q: {\bf R}^m \to {\bf R}$ by $p(x) = x^TJ_mx$, $q(x) = x^TM^TJ_nMx$. Then the set $N
= \{ (p(x),q(x)) \in {\bf R}^2 \,|\, x \in {\bf R}^m \}$ is called the {\sl joint numerical range} of the
matrices $J_m$, $M^TJ_nM$ underlying the quadratic forms $p,q$. It is known \cite{Dines43} that the set $N$
is a convex cone. Lemma \ref{poscond} states the existence of a number $\lambda \geq 0$ such that $q(x) \geq
\lambda p(x)$ for all $x \in {\bf R}^m$. Let $\lambda^*$ be the maximal such $\lambda$. Since $M$ takes the
interior of $L_m$ to the interior of $L_n$, the set $N$ has a non-empty intersection with the open first
orthant. Therefore $\lambda^*$ exists. Moreover, $\lambda^* > 0$, otherwise the matrix $M^TJ_nM$ would be
positive semidefinite, which is not possible if the rank of $M$ is strictly greater than 1. We have $M^TJ_nM
- \lambda^*J_m \succeq 0$ and $M^TJ_nM - \lambda^*J_m - \delta J_m \not\succeq 0$ for any $\delta > 0$. Hence
there exists $x \not= 0$ such that $q(x)-\lambda^*p(x) = x^T(M^TJ_nM - \lambda^*J_m)x = 0$ and $p(x) =
x^TJ_mx \geq 0$. Let us distinguish two cases.

\smallskip

1. There exists a point $x^* \in {\bf R}^m$ such that $q(x^*) - \lambda^*p(x^*) = 0$ and $p(x^*) > 0$.

Without restriction of generality we can choose $x^*$ such that $x^* \in int\,L_m$ and $p(x^*) = 1$. Denote
$Mx^*$ by $y^*$. Since $M$ takes $int\,L_m$ to $int\,L_n$, we have $y^* \in int\,L_n$. In fact,
$(y^*)^TJ_ny^* = q(x^*) = \lambda^* > 0$. Let $U_m$, $U_n$ be automorphisms of the cones $L_m$, $L_n$,
respectively, preserving the quadratic forms $J_m$, $J_n$, respectively, such that $U_m x^* = e_0^m \in {\bf
R}^m$ and $U_n y^* = \sqrt{\lambda^*} e_0^n \in {\bf R}^n$. (Here $e_0^m,e_0^n$ are the unit vectors in the
direction of the coordinate $x_0$ in the respective spaces.) Such automorphisms exist since the Lorentz cones
are homogeneous \cite{Vinberg63}.

Define a map $\tilde M = (\lambda^*)^{-1/2}U_nMU_m^{-1}$. By the positivity of $M$ this map is also positive.
We have $\tilde M e_0^m = (\lambda^*)^{-1/2}U_nMU_m^{-1}U_m x^* = (\lambda^*)^{-1/2}U_ny^* = e_0^n$. Since
$x^*$ is contained in the nullspace of the positive semidefinite matrix $M^TJ_nM - \lambda^*J_m$, we have
$M^TJ_nMx^* = M^TJ_ny^*= \lambda^* J_m x^*$. It follows that
\begin{eqnarray*}
\tilde M^T e_0^n &=& \tilde M^T (J_n e_0^n) = (\lambda^*)^{-1/2}U_m^{-T}M^TU_n^T (U_n^{-T}J_nU_n^{-1}) e_0^n
= (\lambda^*)^{-1/2}U_m^{-T}M^T J_n [(\lambda^*)^{-1/2} y^*] \\
&=& (\lambda^*)^{-1} U_m^{-T} [\lambda^* J_m x^*] = U_m^{-T} J_m x^* = J_m U_m x^* = e_0^m.
\end{eqnarray*}
Hence $\tilde M$ is doubly stochastic.

{\it Remark:} A similar statement for cones of maps that take the positive semidefinite cone to the positive semidefinite
cone was proven by Leonid Gurvits \cite{Gurvits0303055}.

\smallskip

2. For any point $x \in int\,L_m$ we have $q(x) > \lambda^* p(x)$.

We noted above that there exists $x^* \not= 0$ such that $q(x^*) = \lambda^* p(x^*)$ and $p(x^*) \geq 0$.
Since $p(x) > 0$ yields $q(x) > \lambda^* p(x)$, we have $p(x^*) = 0$. Without restriction of generality we
can choose $x^*$ such that $x^* \in \partial L_m$. Denote by $L_N$ the nullspace of the positive semidefinite
matrix $M^TJ_nM - \lambda^*J_m$. This linear subspace contains $x^* \in \partial L_m$ and does not intersect
the interior of $L_m$. Hence it lies in the orthogonal complement to the element $J_mx^* \in \partial L_m$.

On the other hand, $(M^TJ_nM - \lambda^*J_m)x^* = 0$ yields for any vector $v \in Ker\,M$ that $v^T(M^TJ_nM - \lambda^*J_m)x^* =
-\lambda^* v^TJ_mx^* = 0$. Hence the kernel of $M$ lies also in the orthogonal complement of $J_mx^*$. Equivalently, $J_mx^*$ lies in the
image of the matrix $M^T$ and there exists a vector $v \in {\bf R}^n$ such that $J_mx^* = M^Tv$.

Let now $\Delta = J_nv(J_mx^*)^T$ and consider the family of maps $M(\alpha) = M + \alpha \Delta$. For any vector $w \in L_N$ and
for any $\alpha$ we have
\[ [M(\alpha)^TJ_nM(\alpha) - \lambda^* J_m]w = [\alpha \Delta^T J_n M + \alpha M^T J_n \Delta + \alpha^2 \Delta^T J_n \Delta]w = 0,
\]
because $\Delta w = J_nv \langle J_mx^*, w \rangle = 0$ and $\Delta^T J_n M w = J_mx^* v^T J_n^TJ_n M w =
J_mx^* \langle J_mx^*, w \rangle = 0$. Hence there exists $\delta > 0$ such that for all $\alpha \in
(-\delta,+\delta)$ the matrix $M(\alpha)^TJ_nM(\alpha) - \lambda^* J_m$ lies in the face of the
positive semidefinite cone generated by the matrix $M^TJ_nM - \lambda^* J_m$.

Let the matrix $M(\alpha)$ be partitioned as
\[ M(\alpha) = \left( \begin{array}{cc} s(\alpha) & h(\alpha) \\ v(\alpha) & A(\alpha) \end{array} \right).
\]
We have $(s(0)\ h(0)) = (s\ h) \in int L_m$, because otherwise the positive map $M^T$ would take the vector $e_0^n \in int\,L_n$ to the
vector $(s\ h)^T \in \partial L_m$, and $M^T$ would have rank 1 by Lemma \ref{rk1}. Therefore there exists $\delta' > 0$ such that
$(s(\alpha)\ h(\alpha))^T \in int\,L_m$ for all $\alpha \in (-\delta',+\delta')$.

Then by Lemma \ref{poscond} the map $M(\alpha)$ is positive for all $\alpha$ with $|\alpha| < \min(\delta,\delta')$ and hence contained
in the cone ${\cal P}$. Since the rank of $\Delta$ equals 1, but the rank of $M$ is strictly greater than 1, the matrices $M,\Delta$ cannot be collinear. It follows that
$M$ does not generate an extreme ray of ${\cal P}$.

\smallskip

We have proven the following

{\corollary Let $M$ be a positive map of rank strictly greater than 1 and let $M$ generate an extreme ray of ${\cal P}$.
Then there exist automorphisms $U_n \in Aut(L_n)$, $U_m \in Aut(L_m)$ such that
$U_nMU_m$ is doubly stochastic. $\Box$ }

Note that for any automorphisms $U_n \in Aut(L_n)$, $U_m \in Aut(L_m)$ the matrix $U_nMU_m$ generates an extreme ray of ${\cal P}$
if and only if $M$ generates an extreme ray of ${\cal P}$. Let us characterize the extreme rays that are generated by doubly stochastic
matrices. From Lemma \ref{poscond} it follows that a doubly stochastic matrix, partitioned as in (\ref{partition}), is positive if and only if $\sigma_{\max}(A) \leq 1$, where
$\sigma_{\max}$ denotes the maximal singular value.

{\lemma \label{dblstochlem} Let $M$ be a doubly stochastic positive map, partitioned as in (\ref{partition}),
and let $M$ generate an extreme ray of ${\cal P}$. Then all singular values of $A$ equal 1. }

{\it Proof.} Let us assume the contrary. Suppose $M$ is doubly stochastic and positive, partitioned as in (\ref {partition}), with $\sigma_{\min}(A) = \hat\sigma < 1$.
Let $A = UDV$ be the singular value decomposition of $A$ and $\sigma_1,\dots,\sigma_{\min(m-1,n-1)}$ its singular values in decreasing order.
Here $U,V$ are orthogonal matrices of appropriate size and $D = diag(\sigma_1,\sigma_2,\dots,\sigma_{\min(m-1,n-1)})$ is a $(n-1)\times(m-1)$ matrix with the singular values of $A$ on its main diagonal,
all other elements being zero. Note that $\sigma_{\min(m-1,n-1)} = \hat\sigma < 1$. Let us define an affine one-parametric family of diagonal $(n-1)\times(m-1)$ matrices
by $D(\alpha) = diag(\sigma_1,\sigma_2,\dots,\sigma_{\min(n,m)-2},\alpha)$. Then the maps
\[ M(\alpha) = \left( \begin{array}{cc} 1 & 0 \\ 0 & UD(\alpha)V \end{array} \right)
\]
are positive and hence belong to ${\cal P}$ for all $\alpha \in [-1,+1]$. Note that $M = M(\hat\sigma)$. Since these matrices are not proportional for different values of $\alpha$, and $\hat\sigma \in (-1,1)$,
the map $M$ does not generate an extreme ray of the cone ${\cal P}$. This proves the lemma. $\Box$

{\lemma Let $M$ be a doubly stochastic positive map, partitioned as in (\ref{partition}), and let all singular values of $A$ equal 1.
Then $M$ generates an extreme ray of ${\cal P}$ if and only if $\min(n,m) > 2$. }

{\it Proof.} Let $M$ be a map satisfying the assumptions of the lemma. Assume also without restriction of generality that
$n \geq m$. Then we have $A^TA = I_{m-1}$. Let us first show that $M$ does not generate an extreme ray if $\min(n,m) = 2$.

If $m = 2$, then the matrix $A$ is a unit length column vector. Consider the two maps
\[ M_1 = \left( \begin{array}{c} 1 \\ A \end{array} \right) ( 1\ 1 ), \qquad M_2 = \left( \begin{array}{c} 1 \\ -A \end{array} \right) ( 1\ -1 ).
\]
These maps are positive by condition (\ref{poscrk1}) and not proportional. Moreover, we have $M = \frac{1}{2}(M_1+M_2)$. Hence $M$ does
not generate an extreme ray of ${\cal P}$.

Suppose now that $n \geq m \geq 3$. Assume there exists an $n \times m$ matrix
\[ M_{\delta} = \left( \begin{array}{cc} 0 & h_{\delta} \\ v_{\delta} & A_{\delta} \end{array} \right)
\]
and a number $\varepsilon > 0$ such that the map $M(\alpha) = M + \alpha M_{\delta}$ is positive for all $\alpha \in (-\varepsilon,+\varepsilon)$.
The assumption that the upper left element of $M_{\delta}$ is zero does not restrict the generality, because this element can be made zero by
adding to $M_{\delta}$ an appropriate multiple of $M$. Let us develop the positivity condition of Lemma \ref{poscond}. We have that $M(\alpha)$ is positive
if and only if $|\alpha||h_{\delta}| \leq 1$ and there exists $\lambda \geq 0$ such that
\begin{eqnarray}
&& M(\alpha)^TJ_nM(\alpha) - \lambda J_m = \left( \begin{array}{cc} 1 & \alpha v_{\delta}^T \\ \alpha h_{\delta}^T & A^T + \alpha A_{\delta}^T \end{array} \right) J_n
\left( \begin{array}{cc} 1 & \alpha h_{\delta} \\ \alpha v_{\delta} & A + \alpha A_{\delta} \end{array} \right) - \lambda J_m \nonumber\\
&=& \left( \begin{array}{cc} 1-\lambda-\alpha^2|v_{\delta}|^2 & \alpha (h_{\delta}-v_{\delta}^TA) - \alpha^2v_{\delta}^TA_{\delta} \\
\alpha (h_{\delta}^T-A^Tv_{\delta}) - \alpha^2A_{\delta}^Tv_{\delta} &
(\lambda - 1)I_{m-1} - \alpha (A_{\delta}^TA + A^TA_{\delta}) + \alpha^2(h_{\delta}^Th_{\delta} - A_{\delta}^TA_{\delta}) \end{array} \right) \succeq 0. \label{commat}
\end{eqnarray}
Here $\lambda$ may depend on $\alpha$. We obtain in particular $-\alpha (A_{\delta}^TA + A^TA_{\delta}) +
\alpha^2(h_{\delta}^Th_{\delta} - A_{\delta}^TA_{\delta}) \succeq (1-\lambda)I_{m-1} \succeq \alpha^2|v_{\delta}|^2 I_{m-1}
\succeq 0$. A necessary condition for this inequality to hold for all $\alpha \in (-\varepsilon,+\varepsilon)$ is that
$A_{\delta}^TA + A^TA_{\delta} = 0$. It follows that $h_{\delta}^Th_{\delta} \succeq A_{\delta}^TA_{\delta} +
|v_{\delta}|^2 I_{m-1}$. The left-hand side of this inequality is a matrix of rank not exceeding 1, while the
right-hand side is positive semidefinite. Hence the rank of the right-hand side cannot exceed 1 too. Since $m-1 \geq
2$, it follows that $v_{\delta} = 0$ and $A_{\delta}$ is of the form $wh_{\delta}$, where $w$ is a column vector of
appropriate size. This yields the inequality $\alpha^2h_{\delta}^Th_{\delta}(1 - |w|^2) \succeq (1-\lambda)I_{m-1}$,
which implies $\lambda \equiv 1$ for a similar reason. But then the upper left element of matrix (\ref{commat}) is
zero. Therefore $\alpha (h_{\delta}-v_{\delta}^TA) - \alpha^2v_{\delta}^TA_{\delta} = \alpha h_{\delta} = 0$ for all
$\alpha \in (-\varepsilon,+\varepsilon)$ and $h_{\delta} = 0$, $A_{\delta} = wh_{\delta} = 0$. This proves that $M$
generates an extreme ray of ${\cal P}$. $\Box$

\smallskip

Combining the results obtained so far, we can characterize the extreme rays of the cone ${\cal P}$ as follows.

{\lemma Let the positive map $M$ be partitioned as in (\ref{partition}) and suppose that it generates an
extreme ray of the cone ${\cal P}$. Then either $M$ is of rank 1, with $|h| = |v| = s$, or there exist
automorphisms $U_m \in Aut(L_m)$, $U_n \in Aut(L_n)$ such that
\[ U_n M U_m = \left( \begin{array}{cc} 1 & 0 \\ 0 & A' \end{array}
\right)
\]
with all singular values of $A'$ equal to $1$.

If $\min(m,n) \geq 3$, then all matrices of the above types generate extreme
rays. If $\min(m,n) = 2$, then only those of them which are of
rank 1 generate extreme rays. $\Box$}

\smallskip

Note that for any pair of non-zero elements $x,y \in \partial L_m$ in the boundary of the Lorentz cone $L_m$ there exists an
automorphism $U_m$ of $L_m$ that takes $x$ to $y$. Further, for any orthogonal matrix $U_{m-1}$ of dimension $(m-1) \times (m-1)$
the $m \times m$ matrix $diag(1,U_{m-1})$ represents an automorphism of $L_m$. This allows us to reduce the extreme rays of ${\cal P}$
to two canonical forms. Define the two positive maps
\[ M_1 = diag\left( \left( \begin{array}{cc} 1 & 1 \\ 1 & 1 \end{array} \right), 0, \dots, 0 \right) = \left( \begin{array}{cc} {\bf 1}_{2 \times 2} & {\bf 0}_{2 \times (m-2)} \\
{\bf 0}_{(n-2) \times 2} & {\bf 0}_{(n-2) \times (m-2)} \end{array} \right),
\]
\begin{equation} \label{standpos}
M_2 = diag(1,\dots,1) = \left\{ \begin{array}{ccl}
\left( \begin{array}{c} I_m \\ {\bf 0}_{(n-m) \times m} \end{array} \right), & \ & n \geq m, \\
( I_n\ {\bf 0}_{n \times (m-n)}), && n < m. \end{array} \right.
\end{equation}
Here ${\bf 1}_{k \times l}, {\bf 0}_{k \times l}$ denote $k \times l$ matrices filled with ones and zeros, respectively.

{\definition We call a positive map $M$ of {\sl Type I} if there exist automorphisms $U_m \in Aut(L_m)$, $U_n \in Aut(L_n)$
such that $U_n M U_m = M_1$. We call $M$ of {\sl Type II} if there exist automorphisms $U_m \in Aut(L_m)$, $U_n \in Aut(L_n)$
such that $U_n M U_m = M_2$. }

\smallskip

We have the following theorem.

{\theorem \label{extreme}
Let the positive map $M$ generate an extreme ray of the cone ${\cal P}$. Then $M$ is either of Type I or of Type II.

If $\min(m,n) \geq 3$, then all matrices of Types I and II generate extreme rays.
If $\min(m,n) = 2$, then only the matrices of Type I generate extreme rays. $\Box$ }

\medskip

The theorem shows that the structure of the cone ${\cal P}$ of positive maps is more complex than the
structure of the Lorentz cones $L_k$, but is still relatively simple. While the Lorentz cone has only one
kind of extreme rays (which are equivalent with respect to the action of the automorphism group), the cone of
positive maps has two kinds. An exception are the cones of $L_2$-to-$L_n$ positive maps. In this case the
extreme rays form two copies of the boundary $\partial L_n$ of the cone $L_n$ which are located in mutually
orthogonal subspaces.

\section{Largest faces of ball-ball separable cones}

In this section we give a description of the largest faces of ball-ball separable cones, departing from the
two families of extreme rays of the cone of positive maps obtained in the previous section.

\smallskip

We call an element $B$ of the space ${\bf R}^m \otimes {\bf R}^n$ {\sl $L_m \otimes L_n$-separable} or just
{\sl separable} if $B$ can be expressed as a finite sum $\sum_{k=1}^N x_k \otimes y_k$ of product elements
such that $x_k \in L_m, y_k \in L_n$ for all $k = 1,\dots,N$. The separable elements form a convex cone in
${\bf R}^m \otimes {\bf R}^n$, the {\sl separable cone}, which will be denoted by $K_{sep}$. This cone is
dual to the cone ${\cal P}$ of positive maps considered in the previous section. For convenience we will
represent the elements of ${\bf R}^m \otimes {\bf R}^n$ as $n \times m$ matrices such that the scalar product
of a linear map $M: {\bf R}^m \to {\bf R}^n$ with an element $B \in {\bf R}^m \otimes {\bf R}^n$ is given by
$\langle B,M \rangle = tr\,(M^TB) = tr\,(B^TM)$. In this representation a product element $x \otimes y$ is
given by the rank 1 matrix $yx^T$.

It is well-known that the largest faces (i.e.\ non-trivial faces that are not an intersection of other, strictly larger faces) 
of a convex cone $K$ have the form $\{x \in K \,|\, \langle x, y
\rangle = 0\}$, where $y$ generates an extreme ray of the dual cone $K^*$ \cite{Rockafellar}. Let us compute the
faces corresponding to the extreme rays of the cone of positive maps ${\cal P}$ described by Theorem
\ref{extreme}. By this theorem, there are two kinds of extreme rays. These generate two kinds of largest
faces of the separable cone.

Let us define two standard faces of the separable cone $K_{sep}$ by
\[ F_1 = \{ B \in K_{sep} \,|\, \langle B, M_1 \rangle = 0 \}, \quad
F_2 = \{ B \in K_{sep} \,|\, \langle B, M_2 \rangle = 0 \},
\]
where $M_1,M_2$ are the positive maps (\ref{standpos}).

{\definition We call a face $F$ of $K_{sep}$ of {\sl Type I} if there exist automorphisms $U_m \in
Aut(L_m)$, $U_n \in Aut(L_n)$ such that $\{ U_n B U_m \,|\, B \in F \} = F_1$. We call a face $F$ of
$K_{sep}$ of {\sl Type II} if there exist automorphisms $U_m \in Aut(L_m)$, $U_n \in Aut(L_n)$ such that $\{
U_n B U_m \,|\, B \in F \} = F_2$ (or $U_nFU_m = F_2$ for short). }

\smallskip

Hence all faces of Type I are affinely isomorphic to $F_1$, while all faces of Type II are affinely
isomorphic to $F_2$. Let us determine the structure of these two sets.

{\prop \label{F1} $F_1$ is affinely isomorphic to the convex conic hull of the union
\[ Z_1 = \left\{ z = (z_0,z_1,\dots,z_{n+m-2})^T \,\Big|\, z_0 = 1,(z_1-1)^2+\sum_{k=2}^{m-1}z_k^2=1,z_m=\cdots=z_{n+m-2}=0
\right\} \cup
\]
\[ \cup \left\{ z = (z_0,z_1,\dots,z_{n+m-2})^T \,\Big|\, z_0 =
1,z_1=\cdots=z_{m-1}=0,(z_m-1)^2+\sum_{k=m+1}^{n+m-2}z_k^2=1 \right\} \subset {\bf R}^{n+m-1}.
\] }

{\it Remark:} Thus a section of the cone $F_1$ is affinely isomorphic to the convex hull of two intersecting spheres $S^{m-1},S^{n-1}$ which are located in orthogonal subspaces.

{\it Proof.} The set $F_1 = \{ B \in K_{sep} \,|\, \langle B, M_1 \rangle = 0 \}$ is given by the convex hull of
those extreme rays of $K_{sep}$ that are orthogonal to $M_1$. The extreme rays of $K_{sep}$ are tensor
products of the extreme rays generating the individual Lorentz cones $L_m,L_n$, i.e.\ generated by elements
of the form
\begin{equation} \label{tensor}
B = \left( \begin{array}{c} 1 \\ h^T \end{array} \right) \otimes \left( \begin{array}{c} 1 \\ v \end{array} \right) = \left( \begin{array}{cc} 1 & h \\ v & vh \end{array} \right),
\end{equation}
where $h \in {\bf R}^{m-1}$ is a row vector, $v \in {\bf R}^{n-1}$ is a column vector with $|h| = |v| = 1$. We have
\[ \langle B, M_1 \rangle = tr \left( \begin{array}{cc} 1 & h \\ v & vh \end{array} \right)^T \left( \begin{array}{cc} 1 & (e_1^{m-1})^T \\
e_1^{n-1} & e_1^{n-1}(e_1^{m-1})^T \end{array} \right) = (1 + \langle v, e_1^{n-1} \rangle)(1 + \langle h,
e_1^{m-1} \rangle).
\]
Here $e_1^k$ is the unit vector in the direction of the first coordinate in the space ${\bf R}^k$. Therefore
$\langle B, M_1 \rangle = 0$ if and only if $v = -e_1^{n-1}$ or $h = -e_1^{m-1}$. Thus we obtain
\begin{equation} \label{charF1}
F_1 = conv \left\{ \left( \begin{array}{c} 1 \\ -e_1^{n-1} \end{array} \right) x^T + y \,(1\
-(e_1^{m-1})^T) \,\Big|\, x \in \partial L_m, y \in \partial L_n \right\}
\end{equation}
\[ = \left\{ \left(
\begin{array}{ccccc} x_0+y_0 & x_1-y_0 & x_2 & \cdots & x_{m-1} \\ -x_0+y_1 & -x_1-y_1 & -x_2 & \cdots & -x_{m-1} \\
y_2 & -y_2 & \\ \vdots & \vdots & & {\bf 0}_{(n-2) \times (m-2)} & \\ y_{n-1} & -y_{n-1} & \end{array}
\right) \,\Big|\, \begin{array}{c} x_0 \geq |(x_1,\dots,x_{m-1})^T|, \\ y_0 \geq |(y_1,\dots,y_{n-1})^T|
\end{array} \right\}.
\]

It is now easily seen that the affine map $f: {\bf R}^{n+m-1} \to {\bf R}^m \otimes {\bf R}^n$ given by
\[ z = \left( \begin{array}{c} z_0 \\ \vdots \\ z_{n+m-2} \end{array} \right) \mapsto \left( \begin{array}{ccccc} z_0 & z_1-1 & z_2 & \cdots & z_{m-1} \\
z_m-1 & 2-z_0-z_1-z_m & -z_2 & \cdots & -z_{m-1} \\
z_{m+1} & -z_{m+1} & \\
\vdots & \vdots & & {\bf 0}_{(n-2) \times (m-2)} & \\ z_{n+m-2} & -z_{n+m-2} & \end{array} \right)
\]
is an affine bijection between the union $Z_1$ and a set of generators of the cone $F_1$. $\Box$

\smallskip

Let us now consider the second kind of largest faces. Denote by $Sym(k)$ the space of real symmetric $k
\times k$ matrices.

{\prop \label{F2} The face $F_2$ is affinely isomorphic to the set
\[ Z_2 = \{ A \in Sym(\min(n,m)) \,|\, A \succeq 0, A_{00} = tr\, A/2 \},
\]
where $A_{00}$ is the upper left element of the matrix $A$. }

{\it Proof.} Let $n \geq m$ without restriction of generality. Then the positive map
\[ \tilde M_2 = \left( \begin{array}{cc} 1 & {\bf 0}_{1 \times (m-1)} \\ {\bf 0}_{(m-1) \times 1} & -I_{m-1} \\ {\bf 0}_{(n-m) \times 1} & {\bf 0}_{(n-m) \times (m-1)} \end{array} \right)
= M_2 \left( \begin{array}{cc}  1 & {\bf 0}_{1 \times (m-1)} \\ {\bf 0}_{(m-1) \times 1} & -I_{m-1} \end{array} \right)
\]
generates an extreme ray of ${\cal P}$ and is of Type II. Instead of the face $F_2$ we will consider the
isomorphic face $\tilde F_2 = \{ B \in K_{sep} \,|\, \langle B, \tilde M_2 \rangle = 0 \}$.

This face is given by the convex hull
of those extreme rays of $K_{sep}$ that are orthogonal to $\tilde M_2$. Let such an extreme ray be generated by the tensor product (\ref{tensor}).
Let the vector $v$ be partitioned in a subvector $v_a$ of dimension $m-1$ and a subvector $v_b$ of dimension $n-m$. We have
\[ \langle B, \tilde M_2 \rangle = tr \left[ \left( \begin{array}{cc} 1 & h \\ v & vh \end{array} \right)^T \tilde M_2 \right] =
tr \left( \begin{array}{cc} 1 & -v_a^T \\ h^T & -h^Tv_a^T \end{array} \right) = 1 - hv_a.
\]
Note that $|h| = 1$, $|v_a| \leq 1$. Therefore $\langle B, \tilde M_2 \rangle = 0$ if and only if $v_a = h^T$ and $v_b = 0$. Thus $\tilde F_2$ is given by the
convex conic hull of the set
\[ \left\{ \left( \begin{array}{cc}
1 & h \\ h^T & h^Th \\
{\bf 0}_{(n-m) \times 1} & {\bf 0}_{(n-m) \times (m-1)}
\end{array} \right) \,\Big|\, |h| = 1 \right\}.
\]
This hull is equal to the set
\[ conv \left\{ \left( \begin{array}{c} A \\ {\bf 0}_{(n-m) \times m} \end{array} \right) \,\Big|\, A \in Sym(m),\, A \succeq 0,\, rk\, A \leq 1,\, A_{00} = tr\,A/2 \right\}
\]
\[ = \left\{ \left( \begin{array}{c} A \\ {\bf 0}_{(n-m) \times m} \end{array} \right) \,\Big|\, A \in Sym(m),\, A \succeq 0,\, A_{00} = tr\,A/2 \right\}.
\]
The last relation is a consequence of the following fact.

{\it If $L$ is an linear subspace of $Sym(k)$ of codimension 1,
then the intersection of $L$ with the cone $S_+(k)$ of PSD matrices in $Sym(k)$ equals the convex conic hull of all rank 1 PSD matrices
which are contained in $L$. }

Indeed, if these two sets do not coincide, then there exists a linear functional $L'$ on $Sym(k)$ that strictly separates some point in $L \cap S_+(k)$ from 
the convex conic hull of all rank 1 PSD matrices in $L$. But this contradicts the convexity of the joint numerical range \cite{Dines43} of the quadratic forms on
${\bf R}^k$ induced by $L$ and $L'$.

This completes the proof. $\Box$

\smallskip

Above characterizations of the standard faces $F_1,F_2$ allow us to characterize all faces of Types I and II.

{\lemma The faces of Type I are parameterized by a pair of vectors $(h,v)$, where $h \in S^{m-2} \subset
{\bf R}^{m-1}$ is a row vector of length 1 and $v \in S^{n-2} \subset {\bf R}^{n-1}$ is a column vector of
length 1. The face $F_I(h,v)$ corresponding to such a pair $(h,v)$ is given by
\[ F_I(h,v) = conv \left\{ \left( \begin{array}{c} 1 \\ v \end{array} \right) x^T
+ y \,(1\ h) \,\Big|\, x \in \partial L_m, y \in \partial L_n \right\}.
\] }

{\it Proof.} Let $F$ be a face of Type I. Then there exist automorphisms $U_n,U_m$ of $L_n,L_m$,
respectively, such that $F = U_n F_1 U_m$. Define the vectors
\[ h' = U_n \left( \begin{array}{c} 1 \\ -e_1^{n-1} \end{array} \right) \in \partial L_n, \quad v' = U_m^T \left( \begin{array}{c} 1 \\ -e_1^{m-1} \end{array} \right) \in \partial L_m.
\]
Let $h = (h')^T/h'_0$, $v = v'/v'_0$ be normalized multiples of $h',v'$. Then description (\ref{charF1}) of
the standard face $F_1$ shows that $F$ has the form $F_I(h,v)$ defined in the theorem.

On the other hand, the generator sets of $F_I(h,v)$, $F_I(h',v')$ are different whenever $(h,v) \not= (h',v')$. Hence $F_I(h,v) \not= F_I(h',v')$ for $(h,v) \not= (h',v')$. $\Box$

\smallskip

{\corollary \label{corr1} Any two faces of Type I have a non-trivial intersection. }

{\it Proof.} Let $F_I(h,v)$, $F_I(h',v')$ be two faces of Type I. Then the elements
\[ \left( \begin{array}{cc} 1 & h \\ v' & v'h \end{array} \right), \left( \begin{array}{cc} 1 & h' \\ v & vh' \end{array} \right) \in K_{sep}
\]
are contained in both $F_I(h,v)$ and $F_I(h',v')$. $\Box$

{\corollary \label{corr2} Any face of Type I has a non-trivial intersection with any face of Type II. }

{\it Proof.} Let $F_I(h,v)$ be a face of Type I, and let $F_{II}$ be a face of Type II. Then there exist
automorphisms $U_n,U_m$ such that $F_{II} = U_n \tilde F_2 U_m$. We assume $n \geq m$ without loss of
generality. Choose $v' \in \partial L_n$ such that $U_n^{-1}(1\ (v')^T)^T(1\ h)U_m^{-1}$ is in 
$\tilde F_2$. Then the element $(1\ (v')^T)^T(1\ h)$ is shared by the faces $F_{II}$ and $F_I(h,v)$.
$\Box$

\smallskip

On the other hand, faces of Type II do not necessarily have a non-trivial intersection.

\smallskip

In this section we have described the largest faces of the separable cone $K_{sep}$. There are two types of
such faces. All faces of one type are equivalent with respect to the action of the automorphism groups of
the underlying Lorentz cones. Any other non-trivial face is an intersection of some largest faces.
The faces of Type I are affinely isomorphic to the convex conic hull of two
spheres $S^{n-2}$,$S^{m-2}$ which intersect each other in one point, but lie in orthogonal subspaces. The
faces of Type II are intersections of the cone of positive semidefinite $\min(n,m) \times \min(n,m)$-matrices
with a linear subspace of codimension 1. Note that the manifold formed by the union of relative interiors of Type I faces has
$(n+m-1) + (n-2) + (m-2) = 2(n+m)-5$ dimensions, whereas the boundary of $K_{sep}$ is $nm-1$-dimensional. But
$2(n+m)-5 < nm-1$ if $\min(n,m) \geq 3$. Hence the boundary of the $L_2 \otimes L_n$-separable cones is
formed by Type I faces, while the boundary of $K_{sep}$ for $\min(n,m) \geq 3$ is formed by Type II faces.

\section{Radii of largest separable balls}

Extreme rays and largest faces remain invariant under linear bijections. Therefore the results obtained in
the last two sections are extendible to cones that are separable with respect to linear images of standard
Lorentz cones. In this section we compute radii of largest separable balls. These radii are naturally
invariant only under orthogonal mappings, therefore results obtained for $L_m \otimes L_m$-separable cones
will not extend to arbitrary linear images of the Lorentz cones. In order to cover this more general case, we
will consider general ellipsoid-generated cones. We shall compute the radius of the maximal
ellipsoid-ellipsoid separable ball around the tensor product of elements generating the central rays of the
two individual ellipsoid-generated cones. Let us first give a precise definition of an ellipsoid-generated
cone and its central ray.

Let $B \subset E$ be a closed solid ellipsoid with nonempty interior in some $n$-dimensional real vector
space. Suppose that the origin of the space is not contained in $B$. Then the conic hull of $B$ is the image
of the standard Lorentz cone $L_n$ under a regular linear mapping. Moreover, by a rotation it can be
transformed to some standardized ellipsoidal cone
\[
K_{st}(P) = \left\{ (x_0,x_1,\dots,x_{n-1})^T \,\big|\, x_0 \geq \sqrt{x^T P x},\ x = (x_1,\dots,x_{n-1})^T
\right\},
\]
where $P$ is a positive definite symmetric $(n-1)\times(n-1)$-matrix. The set of positive definite symmetric
$(n-1)\times(n-1)$-matrices parameterizes the set of standardized ellipsoidal cones in ${\bf R}^n$. We define
the central ray of $K_{st}(P)$ as the ray generated by the unit vector $e_0 = (1,0,\dots,0)^T$.

Let now $K_1 \subset {\bf R}^m$, $K_2 \subset {\bf R}^n$ be standardized ellipsoidal cones given by positive
definite matrices $P_1,P_2$ of dimensions $(m-1)\times(m-1)$ and $(n-1)\times(n-1)$, respectively:
\begin{eqnarray*}
K_1 &=& K_{st}(P_1) = \left\{ (x_0,x_1,\dots,x_{m-1})^T \in {\bf R}^m \,\big|\, x_0 \geq \sqrt{x^T P_1 x},\ x = (x_1,\dots,x_{m-1})^T \right\}, \\
K_2 &=& K_{st}(P_2) = \left\{ (y_0,y_1,\dots,y_{n-1})^T \in {\bf R}^n \,\big|\, y_0 \geq \sqrt{y^T P_2 y},\ y
= (y_1,\dots,y_{n-1})^T \right\}.
\end{eqnarray*}
Denote by $e_0^N,e_1^N,\dots,e_{N-1}^N$ the unit vectors along the coordinate axes of the space ${\bf R}^N$.
Then the unit vectors $e_0^m,e_0^n$ define the central rays of the cones $K_1,K_2$.

As in the previous section, we shall call an element $B \in {\bf R}^m \otimes {\bf R}^n$ {\sl $K_1 \otimes
K_2$-separable} or just {\sl separable} if $B$ can be expressed as a finite sum $\sum_{k=1}^N x_k \otimes
y_k$ of product elements such that $x_k \in K_1, y_k \in K_2$ for all $k = 1,\dots,N$. The cone of $K_1
\otimes K_2$-separable elements, the {\sl separable cone}, will be denoted by $K_{sep}$. We will represent
the elements of ${\bf R}^m \otimes {\bf R}^n$ as $n \times m$ matrices such that a product element $x \otimes
y$ is given by the rank 1 matrix $yx^T$. Then the product $e_0^m \otimes e_0^n$ is given by a matrix that has
zero elements everywhere except a 1 in the upper left corner.

We shall compute the radius of the largest ball around the unit length vector $e_0^m \otimes e_0^n \in {\bf
R}^m\otimes {\bf R}^n$ consisting of $K_1 \otimes K_2$-separable elements. A ball $B \subset {\bf R}^m\otimes
{\bf R}^n$ consists of separable elements if and only if the conical hull of $B$ is contained in the
separable cone $K_{sep}$. This conical hull is a ball-generated cone, and as such an ellipsoid-generated
cone.

{\lemma \label{radii_rel} Consider the real vector space ${\bf R}^N$. The cone generated by a ball of radius
$\rho < 1$ around the unit vector $e_0$ equals the standardized ellipsoidal cone $K_{st}(r^{-2}I_{N-1})$ with
$r = \frac{\rho}{\sqrt{1-\rho^2}}$. }

{\it Proof.} By the rotational symmetry of the ball-generated cone it must equal a standardized ellipsoidal
cone $K_{st}(P)$ with the matrix $P$ being proportional to the identity matrix. From the definition of
$K_{st}(r^{-2}I_{N-1})$ it follows that $r$ is the radius of the ball created by the intersection of the cone
$K_{st}(r^{-2}I_{N-1})$ with the hyperplane given by the equation $x_0 = 1$. The relation $r =
\frac{\rho}{\sqrt{1-\rho^2}}$ is a consequence of the similarity of appropriate rectangular triangles formed
in the $e_0$-$e_1$ plane of ${\bf R}^N$. $\Box$

Note that $r = \frac{\rho}{\sqrt{1-\rho^2}}$ is a monotonous function of $\rho \in [0,1)$.

\smallskip

Let us denote the ball-generated cone $K_{st}(r^{-2}I_{nm-1})$ by $K_{ball}(r)$. Identify ${\bf R}^{nm}$ with
${\bf R}^m\otimes {\bf R}^n$ by identifying the basis vectors $e_{kn+l}^{mn}$, $k = 0,\dots,m-1$, $l =
0,\dots,n-1$, with the orthonormal basis of tensor products $e_k^m \otimes e_l^n$. Then the cone
$K_{ball}(r)$ is generated by a ball centered on $e_0^m \otimes e_0^n$.

Let $K_{sep}^*,K_{ball}(r)^*$ denote the cones dual to $K_{sep},K_{ball}(r)$. These cones reside in the space
$({\bf R}^{nm})^* = {\bf R}^{nm}$, whose elements will likewise be represented by $n \times m$ matrices. The
scalar product of a matrix $B \in {\bf R}^m\otimes {\bf R}^n$ with a matrix $M \in ({\bf R}^{nm})^*$ will be
defined as $\langle B,M \rangle = tr\,(M^TB) = tr\,(B^TM)$.

{\lemma Let $r$ be the largest number such that the inclusion $K_{ball}(r) \subset K_{sep}$ holds. Then
\begin{equation} \label{maxi}
r^{-1} = \sqrt{ \max \left\{ hP_1h^T + v^TP_2v + tr\,(A^TP_2AP_1) \,\left\vert\, \tilde M = \left(
\begin{array}{cc} 1 & h \\ v & A \end{array} \right) \mbox{ is }L_m\mbox{-to-}L_n\mbox{ positive} \right. \right\} }.
\end{equation} }

{\it Proof.} We have $K_{ball}(r) \subset K_{sep}$ if and only if $K_{sep}^* \subset K_{ball}(r)^*$. By means
of standard linear algebra one establishes that $K_{ball}(r)^* = K_{ball}(r^{-1})$, $K_{st}(P)^* =
K_{st}(P^{-1})$ and $K_{sep}^*$ is the cone of $K_1$-to-$K_2^*$, or $K_{st}(P_1)$-to-$K_{st}(P_2^{-1})$
positive maps. It follows that the largest $r$ satisfying the inclusion $K_{ball}(r) \subset K_{sep}$ equals
the inverse of the smallest $R$ such that any $K_{st}(P_1)$-to-$K_{st}(P_2^{-1})$ positive map lies in the
cone $K_{ball}(R)$.

Let us characterize the cone of $K_{st}(P_1)$-to-$K_{st}(P_2^{-1})$ positive maps and the cone $K_{ball}(R)$.
Let $M: {\bf R}^m \to {\bf R}^n$ be a linear map, partitioned as
\begin{equation} \label{part2}
M = \left( \begin{array}{cc} 1 & h \\ v & A \end{array} \right),
\end{equation}
where $h$ is a row vector of length $m-1$, $v$ is a column vector of length $n-1$ and $A$ is a $(n-1) \times
(m-1)$ matrix.

Since for a positive definite matrix $P$ and for any vector $x$ we have $\sqrt{x^TPx} = |P^{1/2}x|$, we can
characterize the cones $K_1,K_2^*$ as follows:
\begin{eqnarray*}
K_1 &=& \left\{ (x_0,x_1,\dots,x_{m-1})^T \,\big|\, (x_0, (x_1,\dots,x_{m-1})P_1^{1/2})^T \in L_m \right\}, \\
K_2^* &=& \left\{ (y_0,y_1,\dots,y_{n-1})^T \,\big|\, (y_0, (y_1,\dots,y_{n-1})P_2^{-1/2})^T \in L_n
\right\}.
\end{eqnarray*}
It follows that the map $diag(1,P_1^{-1/2})$ is an isomorphism between $L_m$ and $K_1$ and the map
$diag(1,P_2^{-1/2})$ an isomorphism between $K_2^*$ and $L_n$. Hence $M$ is $K_1$-to-$K_2^*$ positive if and
only if the map
\[
\tilde M = diag(1,P_2^{-1/2})\cdot M\cdot diag(1,P_1^{-1/2}) = \left( \begin{array}{cc} 1 & \tilde h \\
\tilde v & \tilde A
\end{array} \right) = \left( \begin{array}{cc} 1 & hP_1^{-1/2} \\ P_2^{-1/2}v & P_2^{-1/2}AP_1^{-1/2}
\end{array} \right)
\]
is $L_m$-to-$L_n$ positive.

Let us examine the cone $K_{ball}(R)$. By definition $M$ is in $K_{ball}(R)$ if $M_{00} \geq
\sqrt{\sum_{(k,l) \not= (0,0)} M_{kl}^2}/R$ (here $M_{kl}$ denotes the elements of $M$). If $M$ is
partitioned as in (\ref{part2}), then $M_{00} = 1$ and $M \in K_{ball}(R)$ if and only if $R \geq
\sqrt{\sum_{(k,l) \not= (0,0)} M_{kl}^2} = \sqrt{|h|^2 + |v|^2 + ||A||_2^2}$.

Hence we obtain the following characterization of the largest number $r$ such that any $K_1$-to-$K_2^*$
positive map is contained in $K_{ball}(r^{-1})$:
\begin{eqnarray*}
r^{-1} &=& \max \left\{ \sqrt{|h|^2 + |v|^2 + ||A||_2^2} \,\left\vert\, \tilde M = \left( \begin{array}{cc} 1
& hP_1^{-1/2} \\ P_2^{-1/2}v & P_2^{-1/2}AP_1^{-1/2} \end{array} \right) \mbox{ is }L_m\mbox{-to-}L_n\mbox{
positive} \right. \right\} \\ &=& \max \left\{ \sqrt{|hP_1^{1/2}|^2 + |P_2^{1/2}v|^2 +
||P_2^{1/2}AP_1^{1/2}||_2^2} \,\left\vert\, \tilde M = \left(
\begin{array}{cc} 1 & h \\ v & A \end{array} \right) \mbox{ is }L_m\mbox{-to-}L_n\mbox{ positive} \right. \right\} \\
&=& \sqrt{ \max \left\{ hP_1h^T + v^TP_2v + tr\,(A^TP_2AP_1) \,\left\vert\, \tilde M = \left(
\begin{array}{cc} 1 & h \\ v & A \end{array} \right) \mbox{ is }L_m\mbox{-to-}L_n\mbox{ positive} \right. \right\}
}.\quad \Box
\end{eqnarray*}

We shall now calculate expression (\ref{maxi}). We have to compute the maximum of the function $F(M) =
hP_1h^T + v^TP_2v + tr\,(A^TP_2AP_1)$ over the set $S$ of $L_m$-to-$L_n$ positive maps $M$ which are
partitioned as in (\ref{part2}), i.e.\ with the upper left element being equal to 1. We shall show that $F$
achieves its maximum either at a rank 1 map or at a doubly stochastic map.

The following lemma is verified by direct calculation.

{\lemma For any integer $n \geq 2$ and any row vector $b \in {\bf R}^{n-1}$ the linear transformation
\[ U_n(b) = \left( \begin{array}{cc} 1 + |b|^2 & \frac{b\left( I_{n-1} - \frac{b^Tb}{(1 + |b|^2)^2} \right)^{-1/2}}{1 + |b|^2} \\
b^T & \left( I_{n-1} - \frac{b^Tb}{(1 + |b|^2)^2} \right)^{-1/2} \end{array} \right)
\]
preserves the quadratic form $J_n$, i.e.\ $J_n = U_n(b)^T J_n U_n(b)$, and is hence an automorphism of the cone $L_n$.
$\Box$ }

\smallskip

Let $M$ be an $L_m$-to-$L_n$ positive map, partitioned as in (\ref{part2}). By the preceding lemma the maps
$U_n(b_n)M$, $MU_m(b_m)$ are also positive for all $b_n \in {\bf R}^{n-1}$, $b_m \in {\bf R}^{m-1}$. The
upper left elements of these products are given by
\[ m_l(b_n) = 1 + |b_n|^2 + \frac{b_n\left( I_{n-1} - \frac{b_n^Tb_n}{(1 + |b_n|^2)^2} \right)^{-1/2}}{1 + |b_n|^2}v > 0, \quad m_r(b_m) = 1 + |b_m|^2 + hb_m^T > 0.
\]
Consider the families of positive maps $M_l(b_n) = U_n(b_n)M/m_l(b_n)$, $M_r(b_m) = MU_m(b_m)/m_r(b_m)$,
parameterized by row vectors $b_n \in {\bf R}^{n-1}$, $b_m \in {\bf R}^{m-1}$. The upper left element of the
corresponding matrices equals 1. Hence $M_l(b_n)$, $M_r(b_m)$ can be partitioned as in (\ref{part2}):
\begin{equation} \label{Ms}
M_l(b_n) = \left( \begin{array}{cc} 1 & h_l(b_n) \\ v_l(b_n) & A_l(b_n) \end{array} \right), \quad M_r(b_m) = \left( \begin{array}{cc} 1 & h_r(b_m) \\ v_r(b_m) & A_r(b_m) \end{array} \right),
\end{equation}
where $h_l(b_n)$, $v_l(b_n)$, $A_l(b_n)$, $h_r(b_m)$, $v_r(b_m)$, $A_r(b_m)$ are vectors and matrices
depending accordingly on the parameter vectors $b_n,b_m$. Define two scalar functions
\begin{eqnarray*}
F_l(b_n) &=& F(M_l(b_n)) = h_l(b_n)P_1h_l(b_n)^T + v_l(b_n)^TP_2v_l(b_n) + tr\,(A_l(b_n)^TP_2A_l(b_n)P_1), \\
F_r(b_m) &=& F(M_r(b_m)) = h_r(b_m)P_1h_r(b_m)^T + v_r(b_m)^TP_2v_r(b_m) + tr\,(A_r(b_m)^TP_2A_r(b_m)P_1).
\end{eqnarray*}

{\lemma Let a $L_m$-to-$L_n$ positive map $M$, partitioned as in (\ref{part2}), realize the maximum of the
function $F$. Then $M$ generates an extreme ray of the cone ${\cal P}$ of $L_m$-to-$L_n$ positive maps. The
corresponding functions $F_l(b_n)$, $F_r(b_m)$ have global maxima at $b_n = 0$, $b_m = 0$, respectively. As a
consequence, their gradients at $b_n = 0$ and $b_m = 0$ vanish. }

{\it Proof.} The function $F(M) = hP_1h^T + v^TP_2v + tr\,(A^TP_2AP_1)$ is strictly convex on the convex set
$S$. Hence its maximum is achieved at an extreme point of this set. Equivalently, the map realizing the
maximum of $F$ generates an extreme ray of ${\cal P}$.

Let the map $M$ realize the maximum of $F$. Define the families of maps (\ref{Ms}). Now note that
$U_n(0),U_m(0)$ are the identity maps, hence $M_l(0) = M_r(0) = M$. Since the maps $M_l(b_n)$, $M_r(b_m)$ are
in $S$ for all $b_n,b_m$, the functions $F_l(b_n)$, $F_r(b_m)$ attain their global maxima at the origin.
$\Box$

{\lemma \label{maxrk1} Let an $L_m$-to-$L_n$ positive map $M$, partitioned as in (\ref{part2}), realize the
maximum of $F$. If $M$ has rank 1, then this maximum is given by $F_{\max} = -1 + (1 + \lambda_{\max}(P_1))(1
+ \lambda_{\max}(P_2))$ (here $\lambda_{\max}$ denotes the maximal eigenvalue). }

{\it Proof.} Let $M$ satisfy the assumptions of the lemma. If $M$ is of rank 1, then $A = vh$. By the
previous lemma $M$ generates an extreme ray of ${\cal P}$. By Lemma \ref{rk1extr} we then have $|h| = |v| =
1$. On the other hand, any pair of unit length vectors $(h',v')$ defines a positive map of rank 1 via
\[ M(h',v') = \left( \begin{array}{cc} 1 & h' \\ v' & v'h' \end{array} \right).
\]
We have
\begin{eqnarray*}
F(M(h',v')) &=& h'P_1(h')^T + (v')^TP_2v' + (v')^TP_2v' + (v')^TP_2v'h'P_1(h')^T \\
&=& -1 + (1+h'P_1(h')^T)(1+(v')^TP_2v').
\end{eqnarray*}
It follows that
\[
F_{\max} = \max_{|h'|=|v'|=1} F(M(h',v')) = -1 + (1 + \lambda_{\max}(P_1))(1 + \lambda_{\max}(P_2)). \quad
\Box
\]

{\lemma \label{sig1} Let $M$ be an $L_m$-to-$L_n$ positive map, partitioned as in (\ref{part2}). Then
$\sigma_{\max}(A) \leq 1$. }

{\it Proof.} Let $M$ be a map satisfying the assumptions of the lemma and suppose that $\sigma =
\sigma_{\max}(A) > 1$. Then there exist unit length column vectors $u,w$ of appropriate dimensions such that
$\sigma u = Aw$ and hence $\sigma = u^TAw$. Without restriction of generality we can assume that $hw-u^Tv
\leq 0$ (otherwise we multiply $u,w$ by $-1$). Then we have
\[ \left( \begin{array}{c} 1 \\ -u \end{array} \right)^T \left( \begin{array}{cc} 1 & h \\ v & A \end{array}
\right) \left( \begin{array}{c} 1 \\ w \end{array} \right) = 1 + hw - u^Tv - \sigma \leq 1 - \sigma < 0.
\]
But
\[ \left( \begin{array}{c} 1 \\ -u \end{array} \right) \in L_n, \quad \left( \begin{array}{c} 1 \\ w \end{array}
\right)\in L_m, \quad M \left( \begin{array}{c} 1 \\ w \end{array} \right) \in L_n,
\]
the last inclusion being due to the positivity of $M$. Hence the scalar product of two vectors in $L_n$ is
negative, which leads to a contradiction with the self-duality of $L_n$. Thus the assumption
$\sigma_{\max}(A) > 1$ was false, which completes the proof. $\Box$

{\lemma Let an $L_m$-to-$L_n$ positive map $M$, partitioned as in (\ref{part2}), realize the maximum of $F$.
Suppose further that this maximum is strictly greater than the maximum over the rank 1 maps established in
Lemma \ref{maxrk1}. Then $h=v=0$. }

{\it Proof.} Let $M$ satisfy the assumptions of the lemma. We shall now compute the gradients of the functions $F_l(b_n)$, $F_r(b_m)$ at $b_n = 0$, $b_m = 0$. We have
\[ U_n'(b)|_{b = 0} = \left( \begin{array}{cc} 0 & b' \\ (b^T)' & 0 \end{array} \right), \quad m_l'(b)|_{b = 0} = b'v, \quad m_r'(b)|_{b = 0} = h(b^T)'.
\]
Hence we obtain
\[ h_l'(b_n)|_{b_n = 0} = b_n'(A - vh), \quad v_l'(b_n)|_{b_n = 0} = (I_{n-1} - vv^T)(b_n^T)', \quad A_l(b_n)|_{b_n = 0} = (b_n^T)'h - (v^T(b_n^T)')A,
\]
\[ h_r'(b_m)|_{b_m = 0} = b_m'(I_{m-1} - h^Th), \quad v_r'(b_m)|_{b_m = 0} = (A-vh)(b_m^T)', \quad A_r(b_m)|_{b_m = 0} =
vb_m' - A(b_m'h^T).
\]
It follows that
\begin{eqnarray*}
F_l'(b_n)|_{b_n = 0} &=& 2\left( h_l'P_1h^T + (v_l^T)'P_2v + tr\,((A_l^T)'P_2AP_1) \right) \\ &=& 2b_n'\left[
(I_{n-1}+P_2)AP_1h^T + (-(hP_1h^T+v^TP_2v+tr(A^TP_2AP_1))I_{n-1} + P_2)v \right], \\
F_r'(b_m)|_{b_m = 0} &=& 2\left( h_r'P_1h^T + (v_r^T)'P_2v + tr\,((A_r^T)'P_2AP_1) \right) \\ &=& 2b_m'\left[
(I_{m-1}+P_1)A^TP_2v + (-(hP_1h^T+v^TP_2v+tr(A^TP_2AP_1))I_{m-1} + P_1)h^T \right].
\end{eqnarray*}
Since the vanishing of the gradient is a necessary condition of maximality of the functions $F_l,F_r$, we obtain the
equations
\begin{eqnarray*}
(I_{n-1}+P_2)AP_1h^T &=& [(1+hP_1h^T+v^TP_2v+tr(A^TP_2AP_1))I_{n-1} - (I_{n-1}+P_2)]v, \\
(I_{m-1}+P_1)A^TP_2v &=& [(1+hP_1h^T+v^TP_2v+tr(A^TP_2AP_1))I_{m-1} - (I_{m-1}+P_1)]h^T.
\end{eqnarray*}
The maximum of $F$ is given by $F_{\max} = 1+hP_1h^T+v^TP_2v+tr(A^TP_2AP_1)$. It follows that
\begin{equation} \label{Aeqs}
AP_1h^T = [(I_{n-1}+P_2)^{-1}F_{\max} - I_{n-1}]v, \quad A^TP_2v = [(I_{m-1}+P_1)^{-1}F_{\max} - I_{m-1}]h^T.
\end{equation}
By the assumptions of the lemma we have $F_{\max} > (1+\lambda_{\max}(P_1))(1+\lambda_{\max}(P_2))$.
Therefore $(I_{n-1}+P_2)^{-1}F_{\max} - I_{n-1} \succ \lambda_{\max}(P_1)I_{n-1}$ and
$(I_{m-1}+P_1)^{-1}F_{\max} - I_{m-1} \succ \lambda_{\max}(P_2)I_{m-1}$. Hence the matrices on the right-hand
sides of (\ref{Aeqs}) are invertible and $h = 0$ implies $v = 0$ and vice versa. Let us assume that $h \not=
0$ and $v \not= 0$. Taking the norms on both sides of equations (\ref{Aeqs}), we get
\[ ||A||_{\infty} \lambda_{\max}(P_1) |h| \geq |AP_1h^T| = |[(I_{n-1}+P_2)^{-1}F_{\max} - I_{n-1}]v| >
\lambda_{\max}(P_1) |v|, \] \[ ||A||_{\infty} \lambda_{\max}(P_2) |v| \geq |A^TP_2v| =
|[(I_{m-1}+P_1)^{-1}F_{\max} - I_{m-1}]h^T| > \lambda_{\max}(P_2) |h|.
\]
Combining, we obtain $||A||_{\infty} = \sigma_{\max}(A) > 1$, which by Lemma \ref{sig1} leads to a
contradiction with the positivity of $M$. Hence $h = v = 0$, which completes the proof. $\Box$

\smallskip

The lemma implies that if $M$ realizes the maximum of $F$ and has a rank greater than 1, then it must be
doubly stochastic.

{\lemma Let $\lambda_1(P_1),\lambda_2(P_1),\dots,\lambda_{m-1}(P_1)$ and
$\lambda_1(P_2),\lambda_2(P_2),\dots,\lambda_{n-1}(P_2)$ be the eigenvalues of the matrices $P_1,P_2$,
respectively, in decreasing order. Then the maximum of $F$ is given by the expression $F_{\max} = \max\{ -1 +
(1 + \lambda_1(P_1))(1 + \lambda_1(P_2)), \sum_{k=1}^{\min(n,m)-1} \lambda_k(P_1) \lambda_k(P_2) \}$. }

{\it Proof.} We have shown above that the maximum of $F$ is achieved either at a rank 1 map, in which case it
equals $F_{\max} = -1 + (1 + \lambda_1(P_1))(1 + \lambda_1(P_2))$, or at a doubly stochastic map. Suppose we
are in the second case, and the map realizing the maximum of $F$ is partitioned as in (\ref{part2}) with
$h=v=0$.

Since the maximum is achieved at a map generating an extreme ray of the cone ${\cal P}$, all singular values
of the matrix $A$ equal 1 by Lemma \ref{dblstochlem}. Assume without restriction of generality that $n \geq
m$. Then the singular value decomposition of $A$ is given by
\[ A = UDV = U \left( \begin{array}{c} I_{m-1} \\ {\bf 0}_{(n-m)\times(m-1)} \end{array} \right) V,
\]
where $U,V$ are orthogonal matrices of appropriate dimensions.

On the other hand, by Lemma \ref{poscond} any pair $(U',V')$ of orthogonal matrices of appropriate size
defines a doubly stochastic positive map
\[ M(U',V') = \left( \begin{array}{cc} 1 & 0 \\ 0 & U'DV' \end{array} \right).
\]
Therefore
\begin{eqnarray*}
F_{\max} &=& \max_{U',V'} F(M(U',V')) = \max_{U',V'} tr(V^TD^TU^TP_2UDVP_1) \\
&=& \max_{U',V'} tr(D^T(U^TP_2U)D(VP_1V^T)).
\end{eqnarray*}
The pair $(U,V)$ of orthogonal matrices maximizes the function $F(M(U',V'))$.

Denote $VP_1V^T$ by $\tilde P_1$ and $U^TP_2U$ by $\tilde P_2$. Then the first order maximality condition is
given by the commutation relations $[D^T\tilde P_2D,\tilde P_1] = [D\tilde P_1 D^T, \tilde P_2] = 0$.
Partition the matrix $\tilde P_2$ as
\[ \tilde P_2 = \left( \begin{array}{cc} \tilde P_2^{11} & \tilde P_2^{12} \\ \tilde P_2^{21} & \tilde
P_2^{22} \end{array} \right),
\]
where $\tilde P_2^{11}$ is of size $(m-1) \times (m-1)$. Then above commutation relations imply $[\tilde
P_1,\tilde P_2^{11}] = 0$, $\tilde P_2^{12} = \tilde P_2^{21} = 0$. Let now $W_{m-1}$ be an orthogonal matrix
that simultaneously block-diagonalizes $\tilde P_1$ and $\tilde P_2^{11}$, and let $W_{n-m}$ be an orthogonal
matrix that diagonalizes $\tilde P_2^{22}$. Now note that $F_{\max} = tr(D^T \tilde P_2 D \tilde P_1) =
tr(D^T [diag(W_{m-1},W_{n-m})\,\tilde P_2\,diag(W_{m-1},W_{n-m})^T] D [W_{m-1}\tilde P_1W_{m-1}^T])$. The
products in brackets are diagonal and have the form $(U')^TP_2U'$, $V'P_1(V')^T$ for some orthogonal matrices
$U',V'$. Hence we can assume without loss of generality that $\tilde P_1,\tilde P_2$ are both diagonal.
Therefore there exist pairwise distinct indices $j_1,\dots,j_{m-1} \in \{1,\dots,n-1\}$ such that $F_{\max} =
\sum_{k=1}^{m-1} \lambda_k(P_1)\lambda_{j_k}(P_2)$. Obviously this sum is maximal if $j_k = k$ for all $k$,
and we arrive at the inequality $F_{\max} \leq \sum_{k=1}^{m-1} \lambda_k(P_1)\lambda_k(P_2)$.

On the other hand, there exist orthogonal matrices $U',V'$ such that \\ $(U')^TP_2U' =
diag(\lambda_1(P_2),\lambda_2(P_2),\dots,\lambda_{n-1}(P_2))$, $V'P_1(V')^T =
diag(\lambda_1(P_1),\dots,\lambda_{m-1}(P_1))$. Then we have $F(M(U',V')) = \sum_{k=1}^{m-1}
\lambda_k(P_1)\lambda_k(P_2)$ and $F_{\max} \geq \sum_{k=1}^{m-1} \lambda_k(P_1)\lambda_k(P_2)$. The proof is
complete. $\Box$

\smallskip

We have proven the following

{\corollary Let $r$ be the largest number such that the inclusion $K_{ball}(r) \subset K_{sep}$ holds. Then
\[ r = \left[ \max\left\{ -1 +
(1 + \lambda_1(P_1))(1 + \lambda_1(P_2)), \sum_{k=1}^{\min(n,m)-1} \lambda_k(P_1) \lambda_k(P_2) \right\}
\right]^{-1/2}. \quad \Box
\] }

\smallskip

By Lemma \ref{radii_rel} we now have the following theorem.

{\theorem \label{thrad} The radius of the largest $K_1 \otimes K_2$-separable ball around $e_0^m \otimes
e_0^n$ is given by
\[ \rho = \left[ \max\left\{
(1 + \lambda_1(P_1))(1 + \lambda_1(P_2)), 1 + \sum_{k=1}^{\min(n,m)-1} \lambda_k(P_1) \lambda_k(P_2) \right\}
\right]^{-1/2}. \quad \Box
\] }

{\corollary \label{ballcor} Let $B_1 \subset {\bf R}^m$, $B_2 \subset {\bf R}^n$ be balls of radii
$\rho_1,\rho_2 < 1$ around the unit vectors $e_0^m,e_0^n$, respectively. Let $K_1,K_2$ be the cones generated
by these balls. Then the radius of the largest $K_1 \otimes K_2$-separable ball around the unit vector $e_0^m
\otimes e_0^n \in {\bf R}^{mn}$ equals
\[ \left[ \max\left\{ \rho_1^{-2}\rho_2^{-2}, 1+ (\min(n,m)-1) (\rho_1^{-2}-1)(\rho_2^{-2}-1) \right\} \right]^{-1/2}.
\] }

The corollary is a direct consequence of the preceding theorem and Lemma \ref{radii_rel}.

\section{Application to multi-qubit systems}

In this section we apply the obtained results to compute largest $K_1\otimes K_2$-separable balls of
bipartite matrices around the identity, where the cones $K_1,K_2$ are generated by balls around the
identities in the factor spaces. We provide the exact value of the radius of such largest balls in dependence
on the radii of the original balls and the dimensions of the matrices. These results will be used to compute
lower bounds on the largest separable ball of unnormalized mixed states for multi-qubit systems.

Denote the space of $k \times k$ hermitian matrices by ${\cal H}(k)$. Let $B_{r_1} \subset {\cal H}(m)$,
$B_{r_2} \subset {\cal H}(n)$ be balls of radii $r_1 < \sqrt{m}$, $r_2 < \sqrt{n}$ around the corresponding
identities $I_m,I_n$ and let $K_1,K_2$ be the conic hulls of these balls. We look for the largest ball around
the identity $I_{nm} \in {\cal H}(mn) = {\cal H}(m)\otimes {\cal H}(n)$ which is contained in the cone of
$K_1 \otimes K_2$-separable matrices.

The following corollary is a consequence of Corollary \ref{ballcor} and the fact that the identity in ${\cal
H}(n) \cong {\bf R}^{n^2}$ has norm $\sqrt{n}$.

{\corollary \label{matrixballs} The largest ball around $I_{nm} \in {\cal H}(mn) = {\cal H}(m)\otimes {\cal
H}(n)$ which is contained in the cone of $K_1 \otimes K_2$-separable matrices has radius
\[ r = \min\left( r_1r_2,
\frac{\sqrt{mn}r_1r_2}{\sqrt{(\min(m^2,n^2)-1)(m-r_1^2)(n-r_2^2) + r_1^2r_2^2}} \right). \ \Box
\] }

\bigskip

We see that for large dimensions and small $r_1,r_2$ $r$ is asymptotically equal to
$\frac{r_1r_2}{\min(m,n)}$. This asymptotics was independently found by Leonid Gurvits\footnote{Leonid Gurvits, personal communication}.

\smallskip

Let us use this result to obtain a bound on the radius of the largest separable ball of unnormalized density
matrices for multi-qubit systems. Let $m = 2$, $r_1 = 1$ and set $n(k) = 2^{k-1}$. Define a sequence $\rho_k$
recursively by $\rho_1 = 1$ and
\begin{eqnarray} \label{thhilf}
\rho_k &=& \min\left( r_1\rho_{k-1},
\frac{\sqrt{mn(k)}r_1\rho_{k-1}}{\sqrt{(\min(m^2,n(k)^2)-1)(m-r_1^2)(n(k)-\rho_{k-1}^2) + r_1^2\rho_{k-1}^2}}
\right) \\ && = \min\left( \rho_{k-1}, \frac{\sqrt{2^k}\rho_{k-1}}{\sqrt{3(2^{k-1}-\rho_{k-1}^2) +
\rho_{k-1}^2}} \right) = \frac{\sqrt{2^k}\rho_{k-1}}{\sqrt{3\cdot 2^{k-1} - 2\rho_{k-1}^2}} \nonumber
\end{eqnarray}
for $k \geq 2$. It follows that
\[ \rho_1^{-2} = 1,\quad \rho_k^{-2} = \frac{3}{2}\rho_{k-1}^{-2} - 2^{-k+1}
\]
and we get the explicit expression
\[ \rho_k^{-2} = \frac{1}{3}\left( \frac{3}{2} \right)^k + 2^{-k},\quad \rho_k = \frac{2^{k/2}}{\sqrt{3^{k-1}+1}}.
\]

{\theorem \label{multiqubit} $\rho_k = \frac{2^{k/2}}{\sqrt{3^{k-1}+1}}$ is a lower bound on the radius of the largest separable
ball of unnormalized multi-partite mixed states of a $k$-qubit system around the identity matrix in the space
${\cal H}(2)^{\otimes k}$. }

{\it Proof.} We prove the theorem by induction.

For a one-qubit system $\rho_1 = 1$ is the radius of the largest ball around $I_2$ in the cone ${\cal
H}_+(2)$ of positive semidefinite hermitian $2 \times 2$ matrices. Hence for $k=1$ the bound $\rho_k$ is
exact.

Assume now that the ball $B_{k-1} \subset {\cal H}(2)^{\otimes(k-1)}$ of radius $\rho_{k-1}$ around the
identity matrix $I_{2^{k-1}} \in {\cal H}(2)^{\otimes(k-1)}$ consists of unnormalized separable states of a
$(k-1)$-qubit system. Let us apply Corollary \ref{matrixballs} with $m = 2$ and $r_1 = 1$. Since the cone
${\cal H}_+(2)$ is isometric to the standard Lorentz cone $L_4 = K_{st}(I_3)$, it will be generated by a ball
of radius 1 around $I_2$ and we get $K_1 = {\cal H}_+(2)$. Let further $n = n(k) = 2^{k-1}$ and $r_2 =
\rho_{k-1}$. If we identify the space ${\cal H}(n)$ with the space ${\cal H}(2)^{\otimes(k-1)}$, then the
cone $K_2$ will be generated by $B_{k-1}$.

But then the ball $B_k \subset {\cal H}(2)^{\otimes k}$ of radius $\rho_k$ around the identity matrix
$I_{2^k}$ is ${\cal H}_+(2) \otimes B_{k-1}$-separable by (\ref{thhilf}) and Corollary \ref{matrixballs}.
Thus it is also ${\cal H}_+(2)^{\otimes k}$-separable by the assumption on $B_{k-1}$. $\Box$

\smallskip

{\it Remark:} $\rho_k$ is the best bound one can obtain by tensoring in the spaces ${\cal H}(2)$ successively
and approximating each time the separable cone by the largest ball-generated cone contained therein. This
general approach was proposed and successfully applied by Gurvits and Barnum in \cite{Gurvits0302102}.

{\it Remark:} Since both factor cones in the ${\cal H}_+(2)^{\otimes 2}$-separable cone are isometric to
$L_4$, Corollary \ref{matrixballs} provides the exact result also for $k = 2$. Gurvits and Barnum obtained the exact
result for a general bipartite space in \cite{Gurvits0204159}.

For a 3-qubit system we get a radius of $\sqrt{4/5}$ instead of $\sqrt{8/11}$ and for $n$-qubit systems with
$n \geq 4$ an improvement of over $12.3\%$ with respect to Gurvits' result in \cite{Gurvits0409095}. The new
bounds imply that with standard NMR preparation technique one needs at least 36 qubits to obtain
entanglement, which is a slightly stronger restriction than the one proven by Gurvits and Barnum \cite{Gurvits0409095}.

\section{Conclusion}

In this contribution we dealt with cones consisiting of elements separable with respect to two Lorentz cones.
Such cones are prospective candidates for the approximation of more complex separable cones, such as the
cones of unnormalized separable states of a multi-partite quantum system. The idea of using a Lorentz cone to
approximate one of the factor cones in a bipartite setting and recursively in a multi-partite setting was
introduced by Leonid Gurvits and Howard Barnum in \cite{Gurvits0302102}. Later they obtained asymptotically exact results on the size of largest
separable balls in \cite{Gurvits0409095}.

We considered different aspects of ball-ball separable cones. Theorem \ref{extreme} describes the extreme
rays generating the cone dual to a ball-ball separable cone, i.e.\ a cone of Lorentz-to-Lorentz positive
maps. There are two kinds of such rays, and all rays of one kind are equivalent under the action induced by
the automorphism groups of the individual Lorentz cones. Correspondingly, the ball-ball separable cones
possess two kinds of largest faces. Here by a "largest" face we mean a non-trivial face that is not
the intersection of other, strictly larger faces. The shape of these faces is described in Propositions
\ref{F1} and \ref{F2}. In Corollaries \ref{corr1} and \ref{corr2} we established that the 
largest faces are highly intersecting each other, unlike the largest faces of a single Lorentz cone. In
Theorem \ref{thrad} and Corollary \ref{ballcor} we compute the radius of the largest ball around an element
on the central ray of a ball-ball separable cone that is contained in this cone. This result extends to the
case of balls in ellipsoid-ellipsoid separable cones. Such cones are affinely isomorphic, but not isometric
to a ball-ball separable cone. The extension to ellipsoid-ellipsoid separable cones allows to use more
flexible approximations of individual factor cones by ellipsoidal cones, which in may be more appropriate
than Lorentz cones in some situations.

Finally, we applied the developed theory to the case of a multi-qubit quantum system. Due to the exactness of
our estimates we were able to sharpen previously available bounds on the radii of maximal separable balls
around the uniformly mixed state. Our bounds in Theorem \ref{multiqubit} are about 12\% tighter than the best bounds obtained so far
\cite{Gurvits0409095}.

\end{document}